\documentclass[11pt]{article} 
\usepackage{amsfonts}
\usepackage[T1]{fontenc}
\usepackage{bm}
\usepackage{amsmath}  
\usepackage{graphicx} 
\usepackage{color}

\newcommand{\be}{ \begin{equation}}
\newcommand{\ee}{\end{equation}}  
\begin{document} 
\def\theequation{\arabic{section}.\arabic{equation}} 
\begin{titlepage} 
\title{Cosmological analogies, Lagrangians, and symmetries for 
convective-radiative heat transfer} 

\author{Valerio Faraoni, Farah Atieh, and Steve Dussault\\\\ 
{\small Department of Physics \& Astronomy, Bishop's University}\\
{\small 2600 College Street, Sherbrooke, Qu\'ebec, Canada J1M~1Z7}
 }

\date{} 
\maketitle 
\thispagestyle{empty} 
\vspace*{1truecm} 
\begin{abstract} 

A formal analogy between the Friedmann equation of relativistic cosmology and 
models of convective-radiative cooling/heating of a body (including Newton's, 
Dulong-Petit's, Newton-Stefan's laws, and a generalization) is discussed. The 
analogy highlights Lagrangians, symmetries, and mathematical properties of the 
solutions of these cooling laws.

\end{abstract} 
\vspace*{1truecm}

%
\end{titlepage}


\def\theequation{\arabic{section}.\arabic{equation}}

\section{Introduction}
\label{sec:1}

Newton announced his model of cooling in 1701, with the rate of heat loss 
by convection from a body of area $A$ at temperature $T$ given 
by\footnote{Since we are going to include the Stefan-Boltzmann law in the 
discussion, $T$ is taken to be the absolute temperature, although this is 
not necessary for convective cooling.}
\begin{equation}
\frac{1}{A} \frac{dQ}{dt} = - h \left( T-T_{\infty} \right) \,,
\end{equation}
where $T_\infty$ is the temperature of the surroundings and 
$h$ is a convection coefficient, assumed to be constant. Newton's law of 
cooling has its limitations: it describes the cooling 
process well only for moderate temperature differences $\theta (t) \equiv 
T(t)-T_{\infty}$ and it  works better for forced convection than 
for natural convection (see Ref.~\cite{Sullivan} for a pedagogical 
review). 

By introducing the heat capacity $C$ of the body, the elementary heat 
transferred in a time $dt$ is $dQ=CdT$ and one obtains 	
\begin{equation}
\frac{dT}{dt} = -k \left( T-T_\infty \right) \label{Newtoncooling}
\end{equation}
with $k \equiv A h/C$. It is mathematically convenient 
to use the temperature difference $\theta(t)$, in 
terms of which $\dot{\theta}=-k\theta$. Assuming that the 
surroundings have a much larger heat capacity than the cooling body, 
$T_{\infty}$ is constant and the solution of Eq.~(\ref{Newtoncooling}) is 
then 
\begin{equation}
T(t) = \left( T_0 -T_{\infty}\right)  \mbox{e}^{-kt} + T_\infty \,,
\end{equation}
where $T_0=T(0)$ is the intial temperature. 
The final state is one of thermal equilibrium $T(t) \simeq T_{\infty}$, 
irrespective of the initial condition $T_0$.

Other models of convective cooling are used to describe regimes in which 
the temperature difference $\theta$ is large, including the 1817  
Dulong-Petit law \cite{DulongPetit}
\begin{equation}
\frac{1}{A} \frac{dQ}{dt} = - g (T-T_a)^n \,,
\end{equation}
where the exponent $n$ ranges between $1.25$ and $1.6$ 
\cite{Taylor,Nelkon,Burns,Sullivan}, which translates 
into
\be
\frac{d\theta}{dt} = - g_0 \, \theta^n \label{DulongPetitcooling}
\end{equation}
($g_0 \equiv gA/C$). 
The solution is
\be
\theta(t)= \frac{\theta_0}{ |t-t_0|^{\frac{1}{n-1}} } \,,
\ee
where $\theta_0= \left[ (n-1)g_0\right]^{\frac{1}{1-n}}$ and $t_0$ is an 
integration 
constant. This solution becomes singular  as $t\rightarrow t_0$, hence 
it cannot be extended arbitrarily in the past. 

Cooling by radiation instead is described (with better accuracy) by 
the 
Stefan-Boltzmann law. In practice, convective Newton cooling 
and radiative cooling often occur together, resulting in the Newton-Stefan 
cooling law 
\begin{equation}
\frac{1}{A} \frac{dQ}{dt} = - h \left( T-T_\infty \right) - \epsilon 
\sigma 
\left( T^4 - T^4_{\infty}\right) \,,
\end{equation} 
where $\sigma$ is the Stefan-Boltzmann constant and 
$\epsilon$ is the emissivity. Now the temperature difference evolves 
according to 
\begin{equation}
\frac{d \theta}{dt} =-\left(  \alpha \theta +\beta\theta^2 + \gamma 
\theta^3 + 
\delta  \theta^4 \right) \,,\label{NewtonStefanexact}
\end{equation}	
where 
\begin{eqnarray}
\alpha &=& \frac{A}{C} \left( h + 4\epsilon \sigma T_{\infty}^3\right) 
\,,\\
&&\nonumber\\
\beta &=& \frac{6A}{C} \, \epsilon \, \sigma T^2_{\infty} \,,\\
&&\nonumber\\
\gamma &=& \frac{ 4A\epsilon \sigma T_{\infty}}{C} \,,\\
&&\nonumber\\
\delta &=& \frac{A \epsilon\sigma}{C} \,.
\end{eqnarray}
A physically meaningful approximation occurs when the temperature 
difference $\theta$ is small and the higher order terms in $\theta$ can be 
dropped 
\cite{Sullivan}, reducing the evolution equation to  
\begin{equation}
\frac{d\theta}{dt} =- \left(\alpha \theta + \beta  \theta^2\right) 
\,.\label{approxNewtonStefan}
\end{equation}
This first order equation integrates to 
\begin{equation}
\theta (t) = \frac{\theta_0 \, \mbox{e}^{-\alpha \, t} }{1+ \Gamma 
\theta_0  \left( 1-\mbox{e}^{-\alpha \, t}\right) }\,,\label{exact}
\end{equation}
where $ \Gamma = \beta/\alpha $ and  $\theta_0=\theta(0)$ is the initial 
temperature  
difference 
between the cooling body and its surroundings.	

If one waits long enough, the exact Newton-Stefan cooling approaches the 
approximated model~(\ref{approxNewtonStefan}). In fact, a 
characteristic feature of spontaneous heat transfer is that 
temperature gradients decrease and temperature differences are smoothed 
out as thermal equilibrium is approached (which is the origin of the 
negative sign in the right hand side of the cooling/heating 
model equations). Therefore, $\theta$ is bound to decrease to a size where 
higher 
order terms in the fourth order polynomial in $\theta$ in 
Eq.~(\ref{NewtonStefanexact}) become negligible 
and the evolution is then ruled only by the linear and quadratic terms. 
Waiting 
longer, only Newton cooling described by the linear term remains relevant. 
In other words, the decreasing exponential solution $\theta_0 \,   
\mbox{e}^{-\alpha t}$ of Newton's model is a late-time attractor in the 
phase space of the solutions of the full Newton-Stefan cooling 
equation~(\ref{NewtonStefanexact}), which ultimately justifies the 
approximated 
model~(\ref{approxNewtonStefan}).

There are formal analogies between these various cooling laws and 
spatially homogeneous and isotropic (or 
Friedmann-Lema\^itre-Robertson-Walker, ``FLRW'' in short) cosmology in the 
context of general relativity. We discuss these analogies after a 
brief summary of FLRW cosmology.

\section{FLRW cosmology}
\label{sec:2}

For the reader's convenience, we recall here the basics of FLRW cosmology 
\cite{Wald,Carroll,Liddle,KT}. The most basic assumption is the Copernican 
principle stating that, on average ({\em i.e.}, over scales larger than 300 
megaparsecs) all spatial points and directions are equivalent: there are no 
preferred points and no preferred directions in space. The assumptions of 
spatial homogeneity and isotropy force the spacetime geometry to be the one 
described by the FLRW line element \cite{Wald,Carroll,Liddle} given, in 
comoving polar coordinates $\left(t, r, \vartheta, \varphi \right)$, 
by\footnote{We adopt the notation of Refs.~\cite{Wald,Carroll}, in which units 
are used such that the speed of light is unity, while $G$ is Newton's 
constant.}
\begin{eqnarray}
ds^2 &=& -dt^2 +a^2(t) \left[ \frac{dr^2}{1-Kr^2} +r^2 \left( d\vartheta^2 + 
\sin^2 \vartheta \, d\varphi^2 \right)\right] \,. \nonumber\\
&&\label{eq:10}
\end{eqnarray}
 The {\em scale factor} $a(t)$ describes how  two points of space at 
fixed comoving  coordinate distance $r_0$ recede from each other with the 
cosmic expansion. Their  physical distance at  
time $t$ is $ l(t)=a(t)r_0$, increasing 
if $ \dot{a}>0$. The function 
$a(t)$ maps the history of the cosmic universe, increasing in an 
expanding universe or decreasing in a contracting one. 

The constant 
$K$ ({\em curvature index}) in 
Eq.~(\ref{eq:10}) describes, respectively, a closed universe (in which 
sections of constant time are 3-spheres) if $K>0$; or a universe with flat  
Euclidean spatial 3-sections if $K=0$; or a universe with hyperbolic 
3-dimensional sections of constant time $t$ if $K<0$ 
\cite{Wald,Carroll,Liddle,KT}. These are the only possible FLRW geometries 
and the dynamics is embodied by the evolution of the scale factor 
$a(t)$.

In cosmology, the matter content of the universe, which causes the 
spacetime to curve, is 
usually modelled by a perfect fluid of energy density $\rho(t)$ and 
isotropic pressure $P(t)$ related by some equation of state. Assuming the 
FLRW geometry, the Einstein field equations of general relativity 
reduce to the so-called Einstein-Friedmann equations for the functions 
$a(t), \rho(t)$, and $P(t)$ \cite{Wald,Carroll,Liddle,KT}
\begin{eqnarray}
&&H^2 \equiv \left( \frac{\dot{a}}{a}\right)^2 
=\frac{\Lambda}{3} + \frac{8\pi G}{3} \, \rho 
-\frac{K}{a^2} \,, \label{eq:11}\\
&&\nonumber\\
&&\frac{\ddot{a}}{a}= -\, \frac{4\pi G}{3} \left( \rho +3P \right) 
-\frac{\Lambda}{3} \,, 
\label{eq:12} \\
&&\nonumber\\
&& \dot{\rho}+3H\left(P+\rho \right)=0 \,,\label{eq:13}
\end{eqnarray}
where $\Lambda$ is Einstein's cosmological constant, an overdot denotes 
differentiation with respect to $t$, and 
$H(t)\equiv 
\dot{a}/a$ is the Hubble function \cite{Wald,Carroll,Liddle,KT}. 
Out of these three equations, only two are independent. If any two are 
given, 
the third one can be derived from them. Without loss of 
generality, we take the Friedmann equation~(\ref{eq:11}) and the 
energy conservation equation~(\ref{eq:13}) as primary, with the 
acceleration 
equation~(\ref{eq:12}) following  from them. 

In cosmology it 
is common (although not compulsory) to assume 
that the cosmic fluid satisfies the barotropic equation of state
\be
P=w\rho \label{eq:15}
\ee
where the ``equation of state 
parameter'' $w=$~const.  Then, Eq.~(\ref{eq:13}) integrates to
\be
\rho(a) = \frac{ \rho_0}{ a^{3(w+1)} }  \label{eq:16} 
\ee
(independent of the curvature index $K$), where $\rho_0$ is a 
positive constant 
determined by the initial conditions. In a spatially flat universe, this gives
\be
a(t) = a_0 \left( t-t_0 \right)^{\frac{2}{3(w+1)} }   \;\;\;\;\;\; (w\neq -1).
\ee

We can already see how the analogy with cooling works: exchanging the 
variables $\left( t, \theta(t) \right) \longrightarrow \left( t, a(t) 
\right)$, the cooling equation is rewritten as 
$ \dot{\theta}= \theta  f(\theta)$ and squared, obtaining
\be
\left( \frac{1}{\theta} \, \frac{d\theta}{dt} \right)^2 = f^2(\theta) 
\,.\label{eq:9}
\ee
The analogy carries through provided that a suitable cosmological fluid 
fills the analogous universe. This is obtained by imposing  the second 
independent equation, {\em i.e.}, the  energy conservation equation 
(\ref{eq:13}).
By comparing Eqs.~(\ref{eq:11}) and~(\ref{eq:9}), we see that it 
must be
\be
\frac{8\pi G}{3} \, \frac{ \rho_0}{ a^{3(w+1)} } =f^2(\theta)
\,. \label{eq:14} 
\ee 
This equation is 
satisfied if $f(\theta)$ is a power law or an inverse power law. One can 
also consider mixtures of non-interacting perfect fluids in which the 
energy densities $\rho_{(i)}$ and partial pressures $P_{(i)}$ of the 
individual fluids add up 
according to Dalton's law. In the simpler case described by 
Eq.~(\ref{eq:14}), provided that $f(\theta)$ has a 
suitable form, the 
analogy between cooling/heating and cosmology is established when the 
analogous universe is filled with a perfect fluid with $P=w\rho$.

A special solution is the de Sitter spacetime corresponding to  a 
spatially flat ($K=0$), empty universe sourced by a positive 
cosmological 
constant $\Lambda$, which has equation of state parameter $w=-1$.  The  
Friedmann equation~(\ref{eq:11})  reduces to 
\begin{equation}
H^2 = \frac{\Lambda}{3} = \, \mbox{const.}
\end{equation}	
and the solution is a simple exponential $a(t)=a_0 \, 
\mbox{e}^{\pm H_{(\pm)}t}$ with constant Hubble function 
$H_{(\pm)}= \pm \sqrt{\Lambda/3}$.

In addition to the previous standard textbook material there are other, 
lesser known, aspects of FLRW cosmology in recent technical literature 
that unveil unknown features of the cooling laws.  They are reported in 
the following subsections.

\subsection{Lagrangian and Hamiltonian}

An effective Lagrangian for spatially flat FLRW cosmology (to which we 
will reduce in the following, although generalizations are possible) is
\be
L\left(a, \dot{a} \right) = 3a\dot{a}^2 + 8\pi G a^3 \rho \,, 
\label{FLRWlagrangian}
\ee
where the function $\rho=\rho(a)$ is specified by the choice of a specific 
barotropic  equation of state $P=P(\rho)$ and by the energy conservation 
equation 
$\dot{\rho}+3H(P+\rho) =0$. Then, the 
Lagrangian~(\ref{FLRWlagrangian}) does not depend explicitly on the 
cosmic time $t$ and the corresponding Hamiltonian is conserved:
\be
{\cal H}=\frac{\partial L}{\partial \dot{a}} \, \dot{a} -L=  
3a\dot{a}^2 -8\pi G a^3 \rho = C \,. \label{aaa} 
\ee
The dynamics of general relativity is constrained dynamics: four of the 
field equations (specifically, the time-time and the time-space components 
of the Einstein field equations) are first order constraints 
\cite{Wald,Carroll}. 
In  the case of the FLRW geometry, due to the high degree of symmetry, the 
only constraint is the Hamiltonian 
constraint (time-time component), which imposes that $C=0$ 
and then~(\ref{aaa}) coincides with the Friedmann equation~(\ref{eq:11})  
\cite{Wald}.

\subsection{Symmetries}

When there is a single term in the right hand side of the cooling equation 
for the temperature difference $\theta \equiv T-T_{\infty}$, the analogous 
FLRW universe filled with a single perfect fluid, and ruled by the 
Friedmann equation~(\ref{eq:11}), is spatially flat. In this case, the 
Einstein-Friedmann equations enjoy certain symmetries 
\cite{Chimento,myPLB,Symmetry2019},  which are studied in the cosmological 
literature, 
mostly in relation with solution-generating techniques 
\cite{Marek1,Marek2,ParsonsBarrow,Chimento,myPLB,BarrowPaliathanasis,Symmetry2019,Pailasetal20}. In these symmetry transformations, one 
rescales time $t$, scale factor $a$, or Hubble function $H$ and changes 
barotropic fluid appropriately, leaving the Einstein-Friedmann equations 
invariant.

The first symmetry \cite{myPLB} is
\begin{eqnarray}
a & \rightarrow & \tilde{a} = \frac{1}{a} \,, \label{symm1a}\\
&&\nonumber\\
w & \rightarrow & \tilde{w} = -(w+2) \,; \label{symm1b}
\end{eqnarray}
it changes an expanding into a contracting universe and {\em 
vice-versa}. When translated to heat transfer, this symmetry is 
unphysical because the temperature 
difference $\theta$ must always be decreasing and this transformation 
turns it into an increasing quantity, corresponding to changing 
the sign of the right hand side of the cooling model equations, therefore to a 
negative 
heat capacity---we will not consider it further.

The second symmetry \cite{Symmetry2019} is
\begin{eqnarray}
a &\rightarrow & \bar{a}=a^s \,,\label{symm2a}\\
&&\nonumber\\
dt & \rightarrow & d\bar{t} = s \, a^{ \frac{3(w+1)(s-1)}{2} } dt 
\,,\label{symm2b}\\
&&\nonumber\\
\rho & \rightarrow & \bar{\rho} = a^{ -3(w+1)(s-1)} \rho \,,\label{symm2c}
\end{eqnarray}
where the real number $s\neq 0$ parametrizes the transformation. These 
symmetries form a 
one-parameter commutative group.

The third type of symmetry transformation \cite{Chimento} is
\begin{eqnarray}
\rho & \rightarrow &  \bar{\rho}=\bar{\rho} (\rho) \,,\label{symm3a}\\
&&\nonumber\\ 
H & \rightarrow &  \bar{H}=\sqrt{ \frac{ \bar{\rho}}{\rho}  }  \, \, H 
\,,\label{symm3b}\\
&& \nonumber\\
P & \rightarrow &  \bar{P }=-\bar{\rho} + \sqrt{ \frac{ \rho}{\bar{\rho}} 
}  \, 
\left( P+\rho\right) \, \frac{ d\bar{\rho} }{d\rho} \,, \label{symm3c}  
\end{eqnarray}
where the function $\bar{\rho}(\rho)$ is positive and regular.

\subsection{Solutions in terms of elementary functions and roulettes}

Methods to solve the Einstein-Friedmann 
equations~(\ref{eq:11})-(\ref{eq:13}) analytically and to study their 
phase space qualitatively are reviewed in 
\cite{oldAmJP,AmJP1,SonegoTalamini}. New results are reported in 
Refs.~\cite{Chen0,Chenetal2015a,Chenetal2015b}, including the proof that the 
graphs of all 
solutions of the Friedmann equation~(\ref{eq:11}) are roulettes 
\cite{Chenetal2015b}. A roulette is the trajectory (in two dimensions) of a 
point 
that lies on a curve rolling without slipping on another given curve. The 
most familiar example is probably the cycloid, which is the trajectory of 
a point on the rim of a bycicle wheel as the bycicle moves forward at 
constant speed on a horizontal surface and the wheel rolls (the graph of the 
corresponding 
solution, the scale factor of a $K=1$ FLRW universe filled with dust, 
appears in all cosmology and relativity textbooks 
\cite{Wald,Carroll,Liddle,KT}).

Finally, one wonders under which conditions it is possible to obtain 
analytical solutions of the Einstein-Friedmann equations in terms of 
elementary functions.  For many choices of the cosmic matter filling the 
universe, by taking the square root of 
the Friedmann equation one obtains 
\be
\frac{da}{dt} = a F(a)  
\ee
where $F(a)$ is a polynomial or a combination of powers. In many cases of 
physical interest, its  
integration is reduced to the task of computing an integral of the form 
\cite{Chen0} 
\begin{eqnarray} \label{I}
I\left(x; p, q, r \right)= \int dx \, x^p \left( a+b \, x^r 
\right)^q     \,,\\
p,q,r \in \mathbb{Q} \,, r\neq 0 \nonumber 
\end{eqnarray}
(if $r=0$ the integral is trivial). This admits the hypergeometric 
function representation
\begin{eqnarray}
I&=& \frac{a^q x^{p+1} }{p+1}   \, {}_2F_1\left( -q, 
\frac{p+1}{r}, \frac{p+r+1}{r}, -\frac{bx^r}{a} \right) 
+\mbox{const.}\nonumber\\
&& 
\end{eqnarray}
which is, however, inconvenient for practical purposes.  A 
necessary and sufficient condition for the  integral~(\ref{I}) to be 
expressed in terms of elementary functions is the Chebysev theorem 
\cite{Chebysev,MarchisottoZakeri}:\\
{\em the integral~(\ref{I}) admits a representation in 
terms of elementary functions if and only if at least one 
of} 
$$ 
\frac{p+1}{r} \,, \;\;\;\; q \,, \;\;\;\; \frac{p+1}{r} + q 
$$
{\em is an integer.}

This theorem will be useful in the following.

\section{Cosmic analogues of cooling laws, Lagrangians, and symmetries}
\label{sec:3}

We now proceed to discuss the cosmological analogues of the various 
first order cooling laws of Sec.~\ref{sec:1}. The analogy provides 
Lagrangians $L=L\left( \theta, \dot{\theta} \right)$, symmetries, and 
mathematical properties of the solutions $\theta(t)$. The 
corresponding Euler-Lagrange 
equation 
\be
\frac{d}{dt} \left( \frac{\partial L}{\partial \dot{\theta}} \right)
-\frac{\partial L}{\partial \theta}=0
\ee
admits a first 
integral which contains an arbitrary integration constant. Only one value 
of this constant reproduces the original cooling equation, as is common in 
the inverse variational problem of finding an action  for 
a first order equation (see \cite{focusnature} for a detailed discussion).

\subsection{Newton cooling}
	
Dividing Newton's law of cooling~(\ref{Newtoncooling}) by $\theta$ and 
squaring yields
\be
\left( \frac{1}{\theta} \, \frac{d\theta}{dt} \right)^2 = k^2 
\,,
\ee
which is formally the same as the Friedmann equation $ H^2= \Lambda/3$ for 
a spatially flat, 
empty  universe with positive cosmological constant $\Lambda=3k^2$. The 
solutions are expanding or contracting de Sitter universes with scale 
factors $
a(t)= a_0 \, \mbox{e}^{\pm kt} $, 
where only the negative sign applies to the analogy for both 
cooling/heating, giving $\theta(t)=\theta_0 \, \mbox{e}^{-kt}$.

Inspired by the FLRW Lagrangian~(\ref{FLRWlagrangian}), one 
finds the effective Lagrangian for Newton's law of cooling
\be
L_1  \left( \theta, \dot{\theta} \right) = \dot{\theta}^2 +\left( \frac{ 
A  h}{C} \right)^2 \theta^2 \,. \label{L1}
\ee
Since $L_1$ does not depend explicitly on time, the corresponding 
Hamiltonian is conserved, ${\cal H}= \dot{\theta}^2 -k^2 \theta^2=C$. 
Choosing  $C=0$ and the negative sign in the square root of the resulting  
equation reproduces Newton's law of cooling.

The symmetry ~(\ref{symm2a})-(\ref{symm2c}) is trivial in this highly 
symmetric situation: it merely 
changes the units of time $t\rightarrow \bar{t}=st$ and rescales the scale 
factor $a\rightarrow \bar{a}=a_0^s \, \mbox{e}^{-k\bar{t}} $, leaving the 
density unchanged, $\bar{\rho}=\rho$.

The other symmetry is also trivial in this case. Since the cosmological 
constant is formally equivalent to a perfect fluid with constant density 
$\rho_{(\Lambda)}=\Lambda/ (8\pi G)$ and pressure 
$P_{(\Lambda)}=-\rho_{(\Lambda)}$, the transformation 
(\ref{symm3a})-(\ref{symm3c}) reduces to the rescaling of $\Lambda$ and of 
the Hubble constant $ \Lambda \rightarrow \bar{\Lambda}= \alpha 
\Lambda $, $ H=\sqrt{\Lambda/3} \rightarrow 
\bar{H}=\sqrt{\bar{\Lambda}/3} = \sqrt{\alpha} H$.

\subsection{Dulong-Petit cooling}

The Dulong-Petit cooling law~(\ref{DulongPetitcooling}) gives
\be
\left( \frac{1}{\theta} \, \frac{d\theta}{dt} \right)^2= g_0^2 \, \theta 
^{2(n-1)} 
\ee
and the analogous Friedmann equation
\be
H^2=\frac{8\pi G}{3} \, \rho 
\ee
describes again a spatially flat analogous universe, but this time the 
cosmological constant is zero and the cosmos is   
filled with a single perfect 
fluid with equation of state parameter 
\be
w= -\frac{(2n+1)}{3} 
\ee
and 
\be
\rho_0= \frac{3g_0^2}{8\pi G} \,.
\ee
This is a phantom ({\em i.e.}, $w<-1$) equation of state and this phantom 
fluid causes the 
universe to end in a Big Rip spacetime singularity at a finite future. 
Phantom fluids \cite{phantom,phantom2} are very exotic forms of dark 
energy often 
invoked to 
explain 
the present acceleration of the cosmic expansion discovered in 1998 with 
type~Ia supernovae \cite{Amendolabook}, but are often favoured by 
cosmological observations. The characteristic feature of a phantom fluid 
is that it makes the 
universe that it fills expand so fast to explode at a finite time in a Big 
Rip singularity. While in the more familiar Big Bang or Big Crunch 
singularities \cite{Wald,Carroll,Liddle,KT} the scale factor $a(t)$ vanishes, 
in a Big Rip it diverges 
instead. The scalar curvature invariants, $\rho$, and $P$ diverge there. 
The expanding and contracting branches on either side of the Big Rip are 
disconnected because the spacetime stops at a curvature singularity, which 
does not belong to the spacetime manifold.

In our case, given the negative sign in the right hand side of the cooling law, 
the analogous 
universe {\em contracts} from a Big Rip at time $t_0$ (where the curvature 
scalars, the energy density $\rho$ 
and the pressure $P$ diverge) with scale factor
\be
a(t)=\frac{a_0}{t-t_0} 
\ee
for $t>t_0$.

In seismology, the Omori-Utsu 
law \cite{Omori,Utsu} giving the time frequency $\dot{n}_s$ of the 
aftershocks following 
a main earthquake shock,
\be 
\dot{n}_{(s)} = -k n_{(s)}^p \,,
\ee
obeys the same equation and has a Big Rip analogy  
\cite{myOmori}. Therefore, there is also an analogy between 
Dulong-Petit cooling and the ``cooling'' of active faults after  a main 
shock. The pictorial expression ``cooling of a seismically active zone'' 
or ``hot zone''  
acquires a precise meaning through this analogy. The dissipation of energy 
through secondary shocks becomes analogous to the removal of heat energy 
from 
a hot body by convection.

The Lagrangian corresponding to Dulong-Petit cooling is 
\be
L_2 \left( \theta, \dot{\theta} \right)= \theta \dot{\theta}^2 +
g_0^2 \, \theta^{2n+1} \,.\label{L2}
\ee
Again, the Hamiltonian ${\cal H}= \theta \dot{\theta}^2 -g_0 ^2 
\theta^{2n+1}$ is conserved and choosing zero value for this ``energy'' 
and the negative sign in the square root reproduces the Dulong-Petit 
cooling law.

Another possible Lagrangian, which is explicitly time-dependent,  is 
\cite{Garra}
\be
L_2 ' = \frac{1}{2} \left( c+t \right)^{\frac{n}{n-1} } \dot{\theta}^2 \,;
\ee
since $\partial L_2'/\partial \theta=0$, the momentum canonically 
conjugated to $\theta$ is conserved,
\be
p_{\theta} \equiv \frac{\partial L_2'}{\partial \dot{\theta} } = 
\left(c+t\right)^{\frac{n}{n-1} } \, \dot{\theta} = c_0 \,,\label{questa}
\ee
where $c_0$ is an integration constant. This equation integrates to
\be
\theta(t)= \frac{-c_0(n-1)}{ (c+t)^{\frac{1}{n-1}}} \,.
\ee
Therefore, 
\be
\frac{1}{(c+t)^{\frac{1}{n-1}} } =\frac{\theta}{c_0 \left(1-n \right)} 
\ee
which, substituted into Eq.~(\ref{questa}), gives
\be
\dot{\theta} = \frac{c_0^{1-n}}{ \left( 1-n \right)^n } \, \theta^n \equiv 
- g_0^2 \, \theta^n \,.
\ee

The symmetry~(\ref{symm2a})-(\ref{symm2c}) gives the new symmetry of the 
Dulong-Petit law
\begin{eqnarray}
\theta &\rightarrow & \bar{\theta}= \theta^s \,,\\
&&\nonumber\\
dt &\rightarrow & d\bar{t} = s \, \theta^{(1-n)(s-1)} dt \,.
\end{eqnarray}
This particular scaling invariance may be useful when scaling 
considerations are needed in applications (for example to scale a  
small system in the lab to industrial size).

Let us consider now the other symmetry~(\ref{symm3a})-(\ref{symm3c}): in order 
to preserve the analogy, the equation of state must be  preserved with 
$\bar{w}=-\left( 2\bar{n}+1\right)/3$, which 
leads to 
\be
\left( \frac{\bar{\rho}}{\rho} \right)^{-3/2} \, \frac{d\bar{\rho}}{d\rho} =
\frac{ \bar{w}+1}{w+1} \,,
\ee
or
\be
d\left( \bar{\rho}^{-1/2} \right) = \frac{ \bar{w}+1}{w+1} \,  
d\left( \rho^{-1/2} \right)\,.
\ee
The solution is 
\be
\bar{\rho} =  \left(\frac{w+1}{\bar{w}+1} \right)^2 \rho = 
\left(\frac{n-1}{\bar{n}-1} \right)^2 \rho \,,
\ee
which implies $ \bar{\rho}_0 = \left(\frac{n-1}{\bar{n}-1} \right)^2 \rho_0$, 
or 
\be 
\bar{g}_0= \frac{n-1}{\bar{n}-1} \, g_0 \,.
\ee 
Equation~(\ref{symm2c}) then gives
\be
\bar{H}=\frac{n-1}{\bar{n}-1} \, H
\ee
and 
\be
\bar{a}(t)=\bar{a}_0 \, t^{ \frac{n-1}{(1-n)(\bar{n}-1)}} = \left[ 
a(t)\right]^{ \frac{ n-1}{\bar{n}-1} } \,.
\ee
Therefore, the transformation for the temperature difference is 
\be
\theta \rightarrow \bar{\theta}=\theta^{ \frac{ n-1}{\bar{n}-1} }\,.
\ee

\subsection{Approximated Newton-Stefan cooling}

Dividing by $\theta$ and squaring, the approximated Newton-Stefan cooling 
law~(\ref{approxNewtonStefan}) becomes
\begin{equation}
 \left( \frac{1}{\theta} \, \frac{d\theta}{dt}\right)^2 = 
\alpha^2  
+ 2 \alpha \beta \, \theta + \beta^2  \theta^2 \,,
\end{equation}
which is analogous to the Friedmann equation~(\ref{eq:11})  
for a universe with flat spatial sections, 
cosmological constant $
\Lambda = 3\alpha^2 $, 
and two perfect fluids with energy densities 
\begin{eqnarray}
\rho_{(1)} &=& \frac{3 \alpha\beta}{4\pi G } \, a \,,\label{rho1}\\
&&\nonumber\\
\rho_{(2)} &=& \frac{3\beta^2}{8\pi G } \, a^2 \,,\label{rho2}
\end{eqnarray}
and equation of state parameters $w_{(1)}= -4/3   \,, w_{(2)}= -5/3    $. 
These are both phantom fluids. Since $\dot{a}<0$ is the only possibility 
in this analogy, they concur with the cosmological constant to 
make the analogous universe contract. The  exact 
solution~(\ref{exact}) is asymptotic to
\be
a(t) \simeq 
\sqrt{\frac{3}{\Lambda}} \,  \, 
\frac{ \theta_0 \, \mbox{e}^{-\sqrt{\Lambda/3} \, t}}{ \sqrt{\Lambda/3} +\beta 
\theta_0}  
\ee
as $t\rightarrow +\infty$, which is explained by noting  that the energy 
densities of the phantom fluids (which scale as the decreasing $a$ or 
$a^2$, respectively) decay, while the energy density of the cosmological 
constant $\rho_{(\Lambda)}=\Lambda/(8\pi G)$ remains constant and comes to 
dominate over the other two as the universe evolves. In other words, the 
contracting de 
Sitter space is an attractor in the phase space of the solutions of the 
approximated Newton-Stefan model. In fact, it is an attractor in the phase 
space of the {\em exact} Newton-Stefan model since higher order terms in 
$\theta$ in the right hand side of Eq.~(\ref{NewtonStefanexact}) become 
negligible sooner than the lower order terms in $\theta$ and $\theta^2$ 
that were retained in the approximation~(\ref{approxNewtonStefan}).

The Lagrangian is now 
\be
L_3 \left( \theta, \dot{\theta} \right) = \theta \dot{\theta}^2 
+ \theta^3 \left( \alpha+ \beta \theta\right)^2 \,.\label{L3}
\ee
The special case $\beta=0$ generates the Lagrangian $L_3 
\Big|_{\beta=0}= \theta L_1$ equivalent to 
(\ref{L1}) for Newton cooling, while setting $\alpha=0$ reproduces 
exactly the 
Lagrangian  (\ref{L2}) for $n=2$ Dulong-Petit cooling.

Since $L_3$ does not depend 
explicitly on the time $t$, the corresponding Hamiltonian is conserved,
\be
{\cal H}=\frac{\partial L_3}{\partial \dot{\theta}} \, \dot{\theta} - L_3 
= \theta \, \dot{\theta}^2 -\theta^3 \left( \alpha +\beta \theta \right)^2 
= C_0 \,,
\ee
where $C_0$ is an integration constant, or 
\be
\left( \frac{1}{\theta} \, \frac{d\theta}{dt} \right)^2 =  
\left( \alpha    + \beta  \theta \right)^2   
+ \frac{C_0}{\theta^3} \,.
\ee
By choosing $C_0=0$, one obtains $\dot{\theta}= \pm \left(\alpha \theta 
+\beta 
\theta^2 \right)$. Only the lower sign reproduces the original problem, as 
discussed in \cite{focusnature}.

Since there are two perfect fluids plus the cosmological constant, the 
symmetries (\ref{symm2a})-(\ref{symm3c}) valid for $K=0$ and a single 
fluid do not apply here. 

If the coefficients are allowed to change signs, the equation describing the 
truncated Newton-Stefan model appears in other 
field of physics and mathematics. For example, the logistic equation can be 
reproduced or, in a simplified laser emission model, the photon 
emission rate $dn/dt$ is related to the number of photons $n(t)$ in an excited 
state by \cite{Haken,Schuch}
\be
\frac{dn}{dt} = -\alpha n -\beta n^2 \,.
\ee

\subsection{More general models}

More general models of convective-radiative cooling, which include the 
exact Newton-Stefan cooling~(\ref{NewtonStefanexact}) are given by
\be
\frac{d\theta}{dt} = - \theta^p \left(\alpha +\beta \theta^r\right)^n 
\,,\label{newmodel}
\ee
where $n$ is a positive integer. The integration of this equation reduces 
to computing the integral
\be
\int d\theta\,  \theta^{-p} \left(\alpha +\beta \theta^r \right)^{-n} \,,
\ee
which is of the form~(\ref{I}) provided that $r$ is rational. In 
practice, since no 
thermal physics experiment is sufficiently precise to distinguish between 
a real number $r$ and its rational approximation, $r$ can always be chosen 
to be rational.  Since here $q$ is the integer $-n$, the Chebysev theorem 
applies and one concludes that the solution of Eq.~(\ref{newmodel}) can 
always be expressed in terms of elementary functions. More precisely, the 
left hand side of the equation
\be
\int d\theta\,  \theta^{-p} \left(\alpha +\beta \theta^r \right)^{-n} 
=-\left( t-t_0 \right) \,,
\ee
where $t_0$ is an integration constant, can be reduced to elementary 
functions. This does not guarantee that this $t(\theta)$  relation can be 
inverted explicitly to provide $\theta(t)$, but in many situations this is 
immaterial. For example, for $p=2, r=1, n=2$, one obtains
\be
\frac{\alpha\beta}{\alpha+\beta \, \theta} +\frac{\alpha}{\theta} +2\beta \ln 
\left( \frac{\theta}{\alpha+\beta\, \theta}\right) =\alpha^3 \left( 
t-t_0\right) \,.
\ee

In all models of cooling/heating for which the cosmological 
analogy is valid, the solution $\theta(t)$ has a graph that is a roulette, 
as is the case for the corresponding analogous universes   
\cite{Chen0}.

\section{Conclusions}
\label{sec:4}

Formal analogies between FLRW cosmology and various models of 
convective-radiative heating/cooling exist. In principle, for an equation 
of the type $\dot{\theta}=\theta f(\theta)$, one can also draw a 
mechanical analogy with the motion of a point particle in one dimension  
subject to an appropriate conservative force (see 
Ref.~\cite{focusnature}), but the cosmic analogy is much more interesting. 

In nature, temperature gradients tend to be smoothed out and disappear 
once final states of thermal equilibrium are reached, unless these 
gradients are maintained by steady heat sources. As is well known, this process 
identifies 
an arrow of time intrinsic to macroscopic objects composed of many 
particles or subsystems, although a preferred  time direction does not 
exist in the equations of fundamental microscopic physics, which are 
time-reversible.  In the analogy between the various phenomenological models of 
thermal 
physics discussed above and the Friedmann equation, the analogue of the 
thermodynamical arrow of time is the formal cosmological arrow of time 
(``formal'' because the real universe expands instead of contracting). All 
comoving objects in these fictitious universes are dragged by the cosmic 
contraction and their proper distances eventually reduce to nothing as the 
scale factor $a(t)$ vanishes asymptotically at late times, just as the 
temperature differences $\theta$ disappear in both convective-radiative 
heating and cooling.

Specifically, we have discussed models of heating / cooling in which the 
function $f^2(\theta)$ appearing in Eq.~(\ref{eq:9}) is a polynomial of 
power laws or inverse power laws. More general forms of $f(\theta)$ can 
also be studied. In this case, imposing the energy conservation equation 
(\ref{eq:13}) leads to a nonlinear barotropic equation of state 
$P=P(\rho)$ for the analogous cosmic fluid. Such equations of state have 
been studied in cosmology, particularly in the last decade in relation 
with hypotetical forms of exotic dark energy 
\cite{BarrowGallowayTipler,ShtanovSahni02,Kofinas03,Calcagni04,Gorini04,Nojirietal05,Barrow04,Diego,Borowiec16}.  
Equations of state of the cosmic fluid of the form $P 
= \sum_{k=1}^m c_k \rho_{(k)}^k$ were studied in Refs. 
\cite{Chenetal2015a,Chenetal2015b,Stefancic05,Framptonetal11,Bouhmadi15}. They 
give rise to a cosmic analogy provided that the energy 
conservation equation is satified, which amounts to
\be
\int \frac{d\rho}{
\sum_{k=1}^m c_k \rho_{(k)}^k +\rho } = -3\ln a \,.
\ee
The special case of a quadratic equation of 
state has been scrutinized more closely 
\cite{NojiriOdintsov04,NojiriOdintsov05,AnandaBruni06a,AnandaBruni06b,Capozzielloetal06,SilvaCosta09}, while pressures depending on fractional powers of 
the density were studied in \cite{Nojirietal05}. Thus far, these equations 
of state are  
completely speculative compared to well established linear ones.

We have shown that, for all models of cooling/heating for which the formal 
analogy with the Friedmann equation holds (including the new models 
(\ref{newmodel})), the solutions are roulettes. Moreover, in all situations of 
practical interest considered here, the analytical solutions can be expressed 
in terms of elementary functions, and new symmetries have been derived. To the 
best of our knowledge, these mathematical properties of the most popular 
cooling models were not discussed in previous literature and emerge due to 
recent results in cosmology \cite{Chen0} and to the analogy with the Friedmann 
equation.

\small 
\section*{Acknowledgments} This work is supported by the Natural 
Sciences \& Engineering Research Council of Canada (Grant no. 2016-03803 
to V.F.) and by Bishop's University.
 

\normalsize


\begin{thebibliography}{99}

\bibitem{Sullivan} C.T. O'Sullivan, {\em Am. J. Phys.} {\bf 58}, 956 
(1990)

\bibitem{DulongPetit} P. Dulong, A. Petit, {\em Ann. Chim. Phys.} {\bf 
7}, 225 (1817)

\bibitem{Taylor} L.W. Taylor, {\em Manual of Advanced Undergraduate 
Experiments in Physics} (Addison-Wesley, London, 1959)

\bibitem{Nelkon} M. Nelkon, P. Parker, {\em Advanced Level Physics} 
(Heinemann, London, 1977)

\bibitem{Burns} D.M. Burns, S.G.G. McDonald, {\em Physics for Biology 
and Premedical Students} (Addison-Wesley, London, 1970)

\bibitem{Wald} R.M. Wald, {\em General Relativity} (Chicago University 
Press, Chicago, 1984)

\bibitem{Carroll} S.M. Carroll, {\em Spacetime and Geometry: An 
Introduction to General Relativity} (Addison Wesley, San Francisco, 2004)

\bibitem{Liddle} A. Liddle, {\em An Introduction to Modern Cosmology} 
(Wiley, New York, 2015)

\bibitem{KT} E.W. Kolb, M.S. Turner, {\em The Early Universe} 
(Addison-Wesley, Redwood City, CA, 1990)

\bibitem{Chimento} L.P. Chimento, {\em Phys. Rev. D} {\bf 65}, 063517 
(2002)

\bibitem{myPLB} V. Faraoni, 
{\em Phys. Lett. B} {\bf 703}, 228 (2011)

\bibitem{Symmetry2019} V. Faraoni, {\em Symmetry} {\bf 12}, 147 (2020)

\bibitem{Pailasetal20} T. Pailas, N. Dimakis, A. Paliathanasis, 
P.A. Terzis,  T. Christodoulakis, 
arXiv:2005.11726

\bibitem{Marek1} M. Szydlowski, M. Heller, {\em Acta Phys. Polon.} {\bf 
B14}, 571 (1983)

\bibitem{Marek2} M. Szydłowski, W. Godlowski, R. Wojtak. {\em Gen. 
Relativ. Gravit.} {\bf 38}, 795 (2006)

\bibitem{ParsonsBarrow} P. Parsons, J.D. Barrow, {\em Class. Quantum 
Grav.} {\bf 12}, 1715 (1995)

\bibitem{BarrowPaliathanasis} J.D. Barrow, A. Paliathanasis, {\em Gen. 
Rel. Gravit.} {\bf 50}, 82 (2018)

\bibitem{oldAmJP} J.E. Felten, R. Isaacman, 
{\em Rev. Mod. Phys.} {\bf 58}, 689 (1986)

\bibitem{AmJP1} V. Faraoni, 
{\em Am. J. Phys.} {\bf 67}, 732 (1999)

\bibitem{SonegoTalamini} S. Sonego, V. Talamini, 
{\em Am. J. Phys.} {\bf 80}, 670 (2012)  

\bibitem{Chen0} S. Chen, G.W. Gibbons, Y. Li, Y. Yang, 
{\em J. Cosmol. Astropart. Phys.} {\bf 1412}, 035 (2014)

\bibitem{Chenetal2015a} A. Chen, G.W. Gibbons, Y. Yang, 
{\em J. Cosmol. Astropart. Phys.} {\bf 2015}, 020 (2015)

\bibitem{Chenetal2015b} S. Chen, G.W. Gibbons, Y. Yang,  
{\em J. Cosmol. Astropart. Phys.} {\bf 10}, 056

\bibitem{Chebysev} M.P. Chebyshev, 
{\em J. Math. Pures Appl.} {\bf  18}, 87 (1853)

\bibitem{MarchisottoZakeri} E.A. Marchisotto, G.-A. Zakeri, 
{\em College Math. J.} {\bf 25}, 295 (1994)

\bibitem{focusnature} V. Faraoni, {\em Eur. J. Phys.}, in press (2020)

\bibitem{phantom} R.R. Caldwell, 
{\em Phys. Lett. B} {\bf 545},  23 (2002)

\bibitem{phantom2} R.R. Caldwell, M. Kamionkowski, N.N. Weinberg, 
{\em Phys. Rev. Lett.} {\bf 91}, 071301 (2003)

\bibitem{Amendolabook} L. Amendola, S. Tsujikawa, {\it Dark Energy, 
Theory and Observations} (Cambridge University Press, Cambridge, England, 
2010)

\bibitem{Omori}F.J. Omori, 
{\em J. Coll. Sci. Imperial Univ. Tokyo} {\bf 7}, 111 (1894)

\bibitem{Utsu} T. Utsu, 
{\em Geophys. Mag.} {\bf 30}, 521 (1961)

\bibitem{myOmori} V. Faraoni, 
{\em Eur. Phys. J. C} {\bf 80}, 445 (2020)  

\bibitem{Garra} R. Garra, private communication.

\bibitem{Haken} H. Haken, {\em Synergetics-An Introduction} (Springer, Berlin, 
1978) 

\bibitem{Schuch} D. Schuch, {\em Quantum Theory from a Nonlinear Perspective},
Fundamental Theories of Physics vol.~191 (Springer, New York, 2018) 

\bibitem{BarrowGallowayTipler} J.D. Barrow, G. Galloway, F.J.T. 
Tipler, 
{\em  Mon. Not. Roy. Astr. Soc.} {\bf 223},  835 (1986)

\bibitem{ShtanovSahni02} Y. Shtanov, V. Sahni,
{\em Class. Quantum Grav.} {\bf 19}, L101-L107 (2002)

\bibitem{Kofinas03} G. Kofinas, R. Maartens, E. Papantonopoulos,  
{\em J. High Energy Phys.} {\bf 2003}, 066 (2003)

\bibitem{Calcagni04} G. Calcagni,
{\em Phys. Rev. D} {\bf 69}, 103508 (2004)

\bibitem{Gorini04}V. Gorini, A. Kamenshchik, U. Moschella, V. 
Pasquier,  
{em Phys. Rev. D} {\bf 69}, 123512 (2004) 

\bibitem{Nojirietal05} S. Nojiri, S.D. Odintsov, S. Tsujikawa,  
{\em Phys. Rev. D} {\bf 71}, 063004 (2005) 

\bibitem{Barrow04} J.D. Barrow, 
{\em Class. Quantum Grav.} {\bf 21}, L79 (2004)

\bibitem{Diego} J. Beltr\'an Jim\'enez, D. Rubiera-Garcia, D. 
S\'aez-G\'omez, V. Salzano, 
{\em Phys. Rev. D} {\bf 94}, 123520 (2016)

\bibitem{Borowiec16} M. Szydlowski, A. Stachowski, A. Borowiec, A. 
Wojnar, 
{\em Eur. Phys. J. C} {\bf 76}, 567 (2016) 

\bibitem{NojiriOdintsov04} S. Nojiri, S.D. Odintsov, 
{\em Phys. Rev. D}, {\bf 70}, 103522 (2004) 

\bibitem{NojiriOdintsov05}
S. Nojiri, S.D. Odintsov,
{\em Phys. Rev. D} {\bf  72}, 023003

\bibitem{Stefancic05}
H. Stefancic, 
{\em Phys. Rev. D} {\bf 71}, 084024 (2005)

\bibitem{Framptonetal11}
P.H. Frampton, K.J. Ludwick, R.J. Scherrer,
{\em Phys.  Rev. D} {\bf  84},  063003 (2011) 

\bibitem{Bouhmadi15}
M. Bouhmadi-L\'opez, A. Errahmani, P. Martin-Moruno, T. Ouali,  
Y. Tavakoli,   
{\em Int. J. Mod. Phys. D} {\bf 24}, 1550078 (2015)  

\bibitem{AnandaBruni06a}
K.N. Ananda, M. Bruni,
{\em Phys. Rev. D} {\bf 74},  023524 (2006)

\bibitem{AnandaBruni06b}
K.N. Ananda, M. Bruni,
{\em Phys. Rev. D} {\bf 74}, 023523 (2006) 

\bibitem{Capozzielloetal06}
S. Capozziello, V.F. Cardone, E. Elizalde, S. Nojiri, S.D. 
Odintsov, 
{\em Phys. Rev. D} {\bf 73}, 043512 (2006) 
 
\bibitem{SilvaCosta09}  S. Silva e Costa, 
{\em Mod. Phys. Lett. A} {\bf 24}, 531 (2009)


\end{thebibliography}
\end{document}